# THE GENERALIZED DENSITY OF STATES IN A ONE-DIMENSIONAL ISING MODEL WITH FERROMAGNETIC AND ANTIFERROMAGNETIC INTERACTIONS

B.V. Kryzhanovsky, V.I. Egorov
*The Kurchatov Institute – Research Institute for System Investigations*
*Moscow, Russian Federation*

**Abstract**. Expressions for the density of states $D(E)$, where $D(E)$ is the number of states of energy $E$, are well known. The present paper offers the expressions for generalized density of states $D_N(E,m)$, where $D_N(E,m)$ is the number of states with energy $E$ and magnetization $m$ in a one-dimensional $N$-spin chain. The expressions obtained here can be considered as reference ones, since all the main characteristics were obtained without them: using the transfer matrix technique or using well-known expressions for the density of states $D(E) = \sum_m D_N(E,m)$. Nevertheless, the knowledge of quantity $D_N(E,m)$ helps to understand the model properties and allows the analysis of the temporal behavior of magnetization $m = m(\tau)$. In particular, we demonstrate that in a one-dimensional model spontaneous magnetization can be observed at a non-zero temperature. However, the spontaneous magnetization can randomly change its sign, which results in the magnetization averaged over a very long observation period becoming zero $\langle m(\tau) \rangle = 0$.

## 1. Introduction

The behavior of a single-dimensional spin system was first studied by Ising in 1925 [1]. The detailed description of the spin system in terms of transfer matrices is given in monograph [2]. The Ising model draws interest as an accurately solvable problem. The single-dimensional model was shown not to have a phase transition at zero temperature. Nevertheless, the model allows us to introduce critical exponents and use the scaling hypothesis and its conclusions [2]. The transfer matrix-based approach made it possible to describe all thermodynamic properties of the single-dimensional model. In particular, the spontaneous magnetization of the model at a non-zero temperature is argued to be strictly zero. However, here the thermodynamic average of magnetization $\langle\langle m \rangle\rangle = 0$ means an ensemble average. The quantum mechanical approach [4-6] does not change these results. It is, however, worth noting that with introduction of long-range interactions into single-dimensional models the phase transitions become possible at non-zero temperatures [7, 8].

The knowledge of quantity $D_N(E,m)$ allows us to monitor the magnetization distribution in the system and discover non-zero spontaneous magnetization at non-zero temperatures. At the same time, the amplitude of spontaneous magnetization is dependent on time $m = m(\tau)$. This time dependence is due to the fact that the energy barrier between two ground system states is so small that the system goes from the vicinity of one ground state with magnetization $m \sim N$ to the vicinity of other ground state with magnetization $m \sim -N$ (where $N$ is the number of states) quite easily. As a result, the time average of spontaneous magnetization $\langle m(\tau) \rangle$ is also zero, in full accordance with the ergodic theorem $\langle m(\tau) \rangle = \langle\langle m \rangle\rangle = 0$. However, it should be noted that the time of the transition between areas with $m \sim N$ and $m \sim -N$ increases noticeably with $N$. It means that the chain starts demonstrating spontaneous magnetization whose sign changes from time to time.

## 2. General expressions

Let us consider a single-dimensional $N$-spin Ising model with only nearest neighbor interactions. The energy of this system is described by the relation:

$$E_H = E - Hm, \quad E = -\sum_{i=1}^{N-1} J_{i,i+1} s_i s_{i+1} - J_{1N} s_1 s_N, \quad m = \sum_{i=1}^{N} s_i. \tag{1}$$

In the relation we picked the term $J_{1N} s_1 s_N$ to be able to consider models with periodic ($J_{1N} \neq 0$) and free ($J_{1N} = 0$) boundary conditions. Below we obtain expressions for generalized density of states $D_N(E,m)$ for both types of the model. We only point out that in the simplest periodic model with interconnections $J_{1N} = J_{i,i+1} = J$ ($i = 1,...,N-1$) the expression for the density of states $D(E) = \sum_m D_N(E,m)$ is well known [3]. Besides, the expression for generalized density of states $D_N(E,m)$ has lately been obtained for this simplest model in [9-10]. A numerical analysis of small-size models ($N = 4,5,6$) and generalization of the result to large $N$ are used in [9-10], while we use simple combinatorics to make analytical conclusions.

First of all, we point out that the knowledge of generalized density of states $D_N(E,m)$, the expression of which is determined below for the case of $H = 0$, is enough to describe the behavior of the system in the presence of a magnetic field ($H \neq 0$). Clear that the number of states with energy $E_H$ and magnetization $m$ is equal to the number of states $D_N(E,m)$ with energy $E$ and magnetization $m$. So, for simplicity in further consideration we assume that $H = 0$.

Before going to determination of $D_N(E,m)$, let us make some general conclusions. To this end let us pass on to new variables $\sigma_i$:

$$\sigma_i = s_i \varepsilon_i, \quad \varepsilon_1 = 1 \text{ and } \varepsilon_i = \text{sgn}(J_{12}J_{23}...J_{i-1,i}), \quad i > 1 \tag{2}$$

Then expression (1) takes ($H = 0$) the form:

$$E = -\sum_{i=1}^{N-1} |J_{i,i+1}| \sigma_i \sigma_{i+1} - \sigma_1 \sigma_N |J_{1N}| \cdot \text{sgn}(J_{12}J_{23}...J_{N-1,N}J_{1N}) \tag{3}$$

We can see that with free boundary conditions ($J_{1,N} = 0$) the energy of a system with alternating interconnections (1) is identical to the energy of a chain with positive interconnections (3). In other words, the problem comes down to the problem of a spin chain with purely ferromagnetic interactions, for which almost all characteristics (partition function, free energy, internal energy, etc.) are known when $|J_{i,i+1}| = 1$.

In the case of a chain with periodic boundary conditions ($J_{1,N} \neq 0$) there are two variants. In the case of an even number of negative interconnections, when $\text{sgn}(J_{12}J_{23}...J_{N-1,N}J_{N1}) = 1$, the problem also comes down to the problem of a positively interconnected chain. Otherwise, it is reduced to the case of a single negative interconnection, which we consider below.

**3. Free boundary conditions ($J_{1N} = 0$)**

Let us consider a free-boundary spin chain whose energy is determined as

$$E = -\sum_{i=1}^{N-1} s_i s_{i+1}. \tag{4}$$

This sort of system has $N$ possible values of energy:

$$E = E_0 + 2k, \quad E_0 = -N + 1, \quad k = 0, 1, ..., (N-1), \tag{5}$$

where $E_0$ is the energy of a doubly degenerate ground state, and the number of states with energy $E = E_0 + 2k$ is determined (Appendix A) by the expression:

$$D(E) = 2 \binom{N-1}{k} \tag{6}$$

Let us now go to the expressions for generalized density of states $D_N(E,m)$. Omitting intermediate calculations (Appendix A), we write down the following results.

For ground states ($k = 0$) we have the obvious expression:

$$D_N(E_0, m) = \begin{cases} 1, & |m| = N \\ 0, & |m| \neq N \end{cases} \tag{7}$$

for states with higher energies ($E = E_0 + 2k$, $k \geq 1$), we have:

$$D_N(E,m) = 2R_k \binom{n-1}{r-1}\binom{N-n-1}{r-1}, \quad R_k = \begin{cases} 1, & \text{odd } k \\ (N-k)/k, & \text{even } k \end{cases} \quad (8)$$

where

$$r = \text{Int}[(k+1)/2], \quad k = (E-E_0)/2, \quad n = (N-m)/2. \quad (9)$$

Here and further $n = (N-m)/2$ is the number of spins inverted relative to the ground state.

Let us make a few general comments about the results of this section.

i) we should note that quantity $D(E) = \sum_m D_N(E,m)$ is identical to expression (6). It is easy to see by performing summation for $m$ (i.e. for $n$) in (8).

ii) states with energy $E = E_0 + 2$ that is closest to the energy of ground states are repeatedly degenerate; the density of states is not zero for all states with magnetization $|m| \leq (N-1)$:

$$D_N(E = E_0 + 2, m) = \begin{cases} 2, & |m| \neq N \\ 0, & |m| = N \end{cases} \quad (10)$$

iii) it follows from (7) – (8) that only states in which the number of inverted spins is related to the energy as $r \leq n \leq N - r$ have a non-zero density, which corresponds to the relationship $||m| \leq (N - 2r)$. Accordingly, for the generalized density of states we can write

$$D_N(E,m) \neq 0 \text{ when } |m| \leq N - 2\,\text{Int}\left(\frac{E+N+1}{4}\right). \quad (11)$$

iv) the maximum of system energy $E_{\max} = N - 1$ is doubly degenerate:

$$D_N(E_{\max}, m) = \begin{cases} 2, & m = 0, \text{ even } N \\ 1, & m = \pm 1, \text{ odd } N \end{cases} \quad (12)$$

As we see, distribution $D_g(E,m)$ on the plane $(E,m)$ looks like a truncated pyramid tapering with $E$. The base of the pyramid is $2N$ wide, the width of its top is 1 or 2 depending on whether $N$ is odd or even.

The most significant conclusion follows from (10): the height of the energy barrier between two ground states is small; there is a chain of $2(N-1)$ doubly degenerate states with energy $E = E_0 + 2$. The chain allows the system to easily migrate from one ground state to another at a none-zero temperature.

**4. Periodic boundary conditions ($J_{1N} = 1$).**

Let us consider a simple spin chain with periodic boundary conditions and the energy described as:

$$E = -s_1 s_N - \sum_{i=1}^{N-1} s_i s_{i+1} \tag{13}$$

This sort of system has $1 + k_{max}$ possible values (see Appendix B):

$$E = E_0 + 4k, \quad E_0 = -N, \quad k = 0, 1, ..., k_{max}, \quad k_{max} = \text{Int}(N/2) \tag{14}$$

where $E_0$ is the energy of the ground state.

First, let's write down the obvious expressions. The ground state is doubly degenerate. Configurations allowing the ground state ($k = 0$) have magnetization $m = N$ and $m = -N$

$$D_N(E_0, m) = \begin{cases} 1, & |m| = N \\ 0, & |m| \neq N \end{cases} \tag{15}$$

The generalized density of high-energy states ($k \geq 1$, $E > E_0$) has the form (see Appendix B):

$$D_N(E, m) = \frac{N}{k}\binom{n-1}{k-1}\binom{N-n-1}{k-1}, \quad k = \frac{E+N}{4}, \quad n = \frac{N-m}{2}. \tag{16}$$

Performing summation over $n$ (i.e. summation over $m$), we get the sought-after result:

$$D(E) = 2\binom{N}{2k} \tag{17}$$

As we mentioned above, the same expression was obtained in [4-5] by numerical analysis of small-size models and extrapolating the result to arbitrary-size models.

Let us comment on the results of this section.

i) It follows from (16) that for the states whose energy $E = E_0 + 4$, is nearest the energy of ground states we have

$$D_N(E_0 + 4, m) = \begin{cases} 0, & |m| = N \\ N, & |m| \neq N \end{cases} \tag{18}$$

It means that there are $(N-1)$ states of energy $E = E_0 + 4$ and magnetization $m$ ($|m| \neq N$), each of which is $N$-fold degenerate.

ii) The amplitude and degeneration degree of the maximum system energy $E_{max}$ depend on the parity of $N$ and can be achieved at different $m$:

$$E_{max} = \begin{cases} N, & \text{even } N \\ N - 2, & \text{odd } N \end{cases} \tag{19}$$

and

$$D_N(E_{max}, m) = \begin{cases} 2, & m = 0, \text{ even } N \\ N, & |m| = 1, \text{ odd } N \end{cases} \tag{20}$$

As we see, with odd $N$ there are two $N$-fold degenerate states with $E = E_{max}$ at $m = \pm 1$.

iii) From (16) it follows that only states in which the number of inverted spins is related to energy as $k \leq n \leq N-k$ (which corresponds to relation $|m| \leq N-2k$) have a non-zero density of states. Accordingly, we have for the density of states:

$$D_N(E,m) \neq 0 \text{ when } |m| \leq \frac{N+E}{2} \tag{21}$$

We see that distribution $D_N(E,m)$ on the plane $(E,m)$ looks like a truncated pyramid tapering with $E$. The base of the pyramid is $2N$ wide, the width of its top is 1 or 2 depending on whether $N$ is odd or even.

As follows from (18), the energy barrier between two ground states is quite low; there is a chain of $N-1$ $N$-fold degenerate states of energy $E = E_0 + 4$. This chain allows the system to easily migrate from one ground state to another at a none-zero temperature.

### 5. A model with ferromagnetic and antiferromagnetic interactions

In previous sections we considered the simplest case of identical interconnections. Let us extrapolate this case to spin chains that have different interconnections, including ferromagnetic and antiferromagnetic interactions.

#### a) Free boundary conditions ($J_{1,N}=0$)

We start with the analysis of a simplest free-boundary system whose energy is described by the expression:

$$E = E_1 + E_2 - J_{12} s_{N_1} s_{N_1+1} \tag{22}$$

where

$$E_1 = -J_{11} \sum_{i=1}^{N_1-1} s_i s_{i+1}, \quad E_2 = -J_{22} \sum_{i=N_1+1}^{N-1} s_i s_{i+1} \tag{23}$$

In other words, the chain of $N$ spins consists of two interacting chains: the first $N_1$-spin chain with interconnections $J_{11} > 0$ and the second $N_2$-spin chain with interconnections $J_{22} > 0$. The two chains interact either in ferromagnetic way ($J_{12} > 0$) or in antiferromagnetic way ($J_{12} < 0$).

We describe the energy of each sub-chain as

$$E_l = J_{ll}(-N_l + 1 + 2k_l), \quad k_l = 1, 2, ..., (N_l - 1), \quad l = 1, 2 \tag{24}$$

and the generalized density of states of each of the sub-systems $D_{N_1}(k_1, n_1)$ and $D_{N_2}(k_2, n_2)$ by expressions (7)-(8) in which we substitute $N_1, k_1, n_1$ or $N_2, k_2, n_2$. Here $n_1$ is the number of inverted spins in the first chain (magnetization $m_1 = N_1 - 2n_1$), and $n_2$ is the number of inverted spins in the second chain (magnetization $m_2 = N_2 - 2n_2$). The total number of inverted spins $n = n_1 + n_2$, and the magnetization of the whole system is

$$m = m_1 + m_2 = N - 2n \tag{25}$$

Correspondingly, the expression for the generalized density of states of the whole system takes (see Appendix C) the form:

$$D_N(E,m) = \sum_{k_1 n_1} \sum_{k_2 n_2} D_{N_1}(k_1, n_1) D_{N_2}(k_2, n_2) \delta_{n, n_1+n_2} \left[ P_{\uparrow\uparrow} \delta_{E, E_1+E_2-J_{12}} + P_{\uparrow\downarrow} \delta_{E, E_1+E_2+J_{12}} \right] \tag{26}$$

Here $\delta_{a,b}$ is the Kronecker symbol, $P_{\uparrow\uparrow}$ is the probability that spins $s_{N_1}$ and $s_{N_1+1}$ have the same direction:

$$P_{\uparrow\uparrow} = \begin{cases} a_1 a_2 + b_1 b_2, & k_1 \text{ and } k_2 \text{ are even} \\ 1/2, & \text{otherwise} \end{cases} \tag{27}$$

and $P_{\uparrow\downarrow}$ is the probability of spins $s_{N_1}$ and $s_{N_1+1}$ having opposite direction:

$$P_{\uparrow\downarrow} = \begin{cases} a_1 b_2 + a_2 b_1, & k_1 \text{ and } k_2 \text{ are even} \\ 1/2, & \text{otherwise} \end{cases} \tag{28}$$

where

$$a_1 = \frac{n_1 - k_1/2}{N_1 - k_1}, \quad b_1 = \frac{N_1 - n_1 - k_1/2}{N_1 - k_1} \tag{29}$$

$$a_2 = \frac{n_2 - k_2/2}{N_2 - k_2}, \quad b_2 = \frac{N_2 - n_2 - k_2/2}{N_2 - k_2} \tag{30}$$

Generalization to the case of a more complex system looks obvious. Let the chain, for instance, consists of three interacting sub-chains of lengths $N_1$, $N_2$, $N_3$ with different interconnection within each of them. In this event the algorithm of determining the generalized density of states is as follows. First, we use the formulae to compute the density of states of the system consisting of two chains $N_1$ and $N_2$. Then we turn to formulae (26) – (30) to make convolution of the expressions for the united two-chain system ($N_1$ and $N_2$) and $N_3$-spin chain.

**b) Periodic boundary conditions ($J_{1,N} \neq 0$)**

Let us consider the simple case of a closed chain whose energy has the form:

$$E = E_1 + E_2 - J_{12}\left( s_1 s_N + s_{N_1} s_{N_1+1} \right) \tag{31}$$

where

$$E_1 = -J_{11} \sum_{i=1}^{N_1-1} s_i s_{i+1}, \quad E_2 = -J_{22} \sum_{i=N_1+1}^{N-1} s_i s_{i+1} \tag{32}$$

In other words, the $N$-spin chain consists of two interacting sub-chains: the first $N_1$-spin chain with interconnections $J_{11} > 0$ and the second $N_2$-spin chain with interconnections $J_{22} > 0$. The two sub-chains interaction is either ferromagnetic ($J_{12} > 0$) or antiferromagnetic ($J_{12} < 0$).

The chain like that can be represented as a sum of two free-boundary chains whose ends are coupled (see Appendix D). We describe the energy of each sub-chain as

$$E_l = J_{ll}\left(-N_l + 1 + 2k_l\right), \quad k_l = 1, 2, \ldots, (N_l - 1), \quad l = 1, 2 \tag{33}$$

and the generalized density of states of each of the sub-systems $D_{N_1}(k_1, n_1)$ and $D_{N_2}(k_2, n_2)$ by expressions (7) – (8) in which we should substitute $N_1, k_1, n_1$ or $N_2, k_2, n_2$. Here $n_1$ is the number of inverted spins in the first chain (partial magnetization $m_1 = N_1 - 2n_1$), and $n_2$ is the number of inverted spins in the second chain (partial magnetization $m_2 = N_2 - 2n_2$). The total number of inverted spins $n$, and the magnetization of the whole system is

$$n = n_1 + n_2, \quad m = m_1 + m_2 = N - 2n \tag{34}$$

Correspondingly, the expression for the generalized density of states of the whole system takes (see Appendix D) the form:

$$D_N(E, m) = \sum_{k_1 n_1} \sum_{k_2 n_2} D_{N_1}(k_1, n_1) D_{N_2}(k_2, n_2) \delta_{n, n_1 + n_2} w_{k_1 k_2} \tag{35}$$

where

$$w_{k_1 k_2} = \begin{cases} \delta_{E, E_1 + E_2}, & \text{if } k_1 \bmod 2 \neq k_2 \bmod 2 \\ P_{\uparrow\uparrow} \delta_{E, E_1 + E_2 - 2J_{12}} + P_{\uparrow\downarrow} \delta_{E, E_1 + E_2 + 2J_{12}}, & \text{if } k_1 \bmod 2 = k_2 \bmod 2 \end{cases} \tag{36}$$

and quantities $P_{\uparrow\uparrow}$ and $P_{\uparrow\downarrow}$ are described by expressions (27) – (30). As is seen, the closed chain does not differ much from the chain with free boundary conditions. The generalization to a more complicated system is done by using the convolution of corresponding expressions (35).

### c) Periodic chain with one negative interconnection ($J_{1,N} < 0$)

In conclusion of this section let us consider the case when the problem of a closed chain with alternating interconnection is not reduced to the problem of a chain with positive interconnection. As follows from (3), it is about a closed chain with an odd number of negative interconnections. For illustration, let us take the simple case (see Appendix E for more general considerations) when the energy of the chain is defined by the expression:

$$E = s_1 s_N - \sum_{i=1}^{N-1} s_i s_{i+1} \tag{37}$$

This sort of system has $1 + k_{max}$ possible values of energy (see Appendix E):

$$E = E_0 + 4k, \quad E_0 = -N + 2, \quad k = 0, 1, ..., k_{max}, \quad k_{max} = \text{Int}\left(\frac{N+1}{2}\right) \tag{38}$$

where $E_0$ is the energy of the ground state. The generalized density of states for this chain (see Appendix E) takes the form:

$$D_N(E, m) = \frac{2n(N-n) - kN}{n(N-n)} \cdot \binom{n}{k}\binom{N-n}{k}, \quad m = N - 2n \tag{39}$$

By summing over $n$ in (39), we get the expression for the density of states:

$$D(E) = \sum_{n=0}^{N} D_N(E, m) = \frac{N - 2k}{2k + 1} \cdot \binom{N}{2k} \tag{40}$$

This expression is somewhat different from the known expression for the case of purely ferromagnetic interactions (17).

It follows from (39) that the ground state is 2$N$-fold degenerate. Correspondingly, the system can easily move between states with $m \sim N$ and $m \sim -N$.

### 6. Magnetization distribution $P(m)$

Let us consider the temperature dependence of the magnetization distribution of the system. The probability of the system being in the state with magnetization $m$ is determined by the formula:

$$P(m) = \frac{1}{Z} \sum_{E} D_N(E, m) e^{-\beta E + \beta m H} \tag{41}$$

where $\beta$ is the inverse temperature, and $Z$ is the normalization constant (partition function):

$$Z = \sum_{m} \sum_{E} D_N(E, m) e^{-\beta E + \beta m H} \tag{42}$$

We consider the general properties of this sort of distribution $P(m)$ and behavior of spontaneous magnetization by using two example models: a model with positive interconnections and a model with a few negative interconnections. Below we will show that for a single-dimensional model it is possible to introduce three important values of inverse temperature $\beta_0 < \beta_1 < \beta_2$. When $0 \leq \beta < \beta_0$, distribution $P(m)$ is a single-mode distribution whose center is at $m = 0$. When $\beta = \beta_0$, two more peaks appear in distribution $P(m)$ at points corresponding to ground-state magnetizations. When $\beta = \beta_1$, the additional peaks become as high as the central peak. When $\beta = \beta_2$, the period of the system staying in the ground states becomes equal to the period of the system being in all other states.

*a) Distribution P(m) in a model with positive interconnections*

Let us consider the properties of a single-dimensional model with periodic boundary conditions whose energy is described by (13). In this event, substituting expressions (15)-(16) in (42) and summing (42) over $E$ and $m$, we obtain well known expressions [1-3]:

$$Z = \lambda_1^N + \lambda_2^N, \quad \lambda_{1,2} = e^\beta \cosh \beta H \pm \left(e^{2\beta} \sinh^2 \beta H + e^{-2\beta}\right)^{\frac{1}{2}}, \qquad (43)$$

which in case of $H = 0$ takes the form:

$$Z = \left(e^\beta + e^{-\beta}\right)^N + \left(e^\beta - e^{-\beta}\right)^N \qquad (44)$$

Let us examine the mode composition of distribution $P(m)$. To this end we perform the graphical analysis of $P(m)$ when $H = 0$. Figure 1 shows how curve $P(m)$ changes with inverse temperature. We see (Fig. 1a) that when the inverse temperature varies within interval $0 \leq \beta \leq \beta_0$ ($\beta_0$ will be determined below), $P(m)$ is a single-mode distribution with the center at point $m = 0$. When $\beta > \beta_0$ (Fig. 1b), $P(m)$ turns into a three-mode distribution: in addition to the central peak at m = 0, two peaks of the same height appear at the end points $m = \pm N$ of the distribution. When $\beta = \beta_1$ these two peaks become as high as the central peak. With increasing $\beta$ (Fig. 1c), the side peaks become dominant, while the height of the central peak approaches zero.

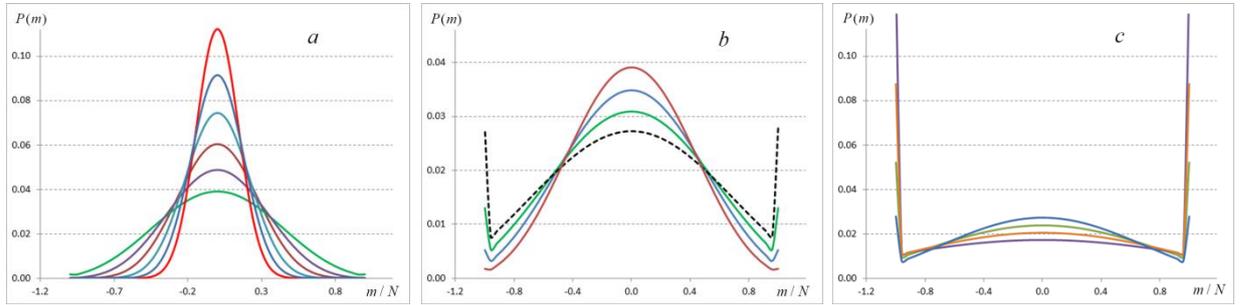

*Figure* 1. Temperature dependence of distribution $P(m)$: a) single-mode distribution for $0 \leq \beta \leq \beta_0$; b) three-mode distribution when $\beta_0 \leq \beta \leq \beta_1$ (the dotted line corresponds to $\beta = \beta_1$); c) interval $\beta \geq \beta_1$.

Let us now define values $\beta_0$ and $\beta_1$. The value $\beta_0$ is determined by the condition of peaks emerging at the end points $m = \pm N$ of distribution P(m). It is easy to notice that the appearance of peaks at points $m = \pm N$ corresponds to condition $P(|m| = N) \geq P(|m| = N-1)$, i.e. condition $(e^{-\beta E_0}/Z) \geq Ne^{-4\beta}(e^{-\beta E_0}/Z)$. Hence it follows that

$$\beta_0 = \frac{1}{4} \operatorname{Ln} N \qquad (45)$$

The value $\beta_1$ is found from condition $P(m=0) = P(|m|=N)$. Let us evaluate $P(m=0)$ in the case of $N \gg 1$. For this purpose, we substitute corresponding expressions for $D_N(E,m)$ into (41) and replace the summation over $k$ (summation over $E = E_0 + 4k$) by the integration with the method of steepest descent. After all reductions the condition takes the form:

$$\sqrt{\frac{2}{\pi N}} e^{-\beta} \left(1 + e^{-2\beta}\right)^N = 1 \tag{46}$$

Hence the evaluative expression:

$$\beta_1 \simeq \frac{1}{2} \text{Ln} N - \frac{1}{2} \text{Ln} \text{Ln} N \tag{47}$$

As follows from the above analysis, when $\beta > \beta_0$, $P(m)$ becomes a three-mode distribution, and when $\beta > \beta_1$, the modes at points $m = \pm N$ become predominant. It would seem that with $\beta > \beta_1$ the system in the thermodynamic equilibrium should pass to states with $|m| \neq 0$, i.e. there should be spontaneous magnetization. The effect really takes place, yet there are some peculiarities that we consider below.

### b) Distribution P(m) in a model with positive and negative interconnections

Let us consider how the presence of negative interconnections in the spin chain changes distribution P(m). We confine ourselves to the simplest case corresponding to expression (26) when $J_{11} = J_{22} = 1$, $J_{12} = -1$. Here we omit expressions for $\beta_0$ and $\beta_1$ because of their complication.

Curves $P(m)$ for different temperatures are shown in Figure 2. As we see, the graphs are heavily dependent on the relations between $N_1$ and $N_2$. As shown in Figure 2a, $P(m)$ remains a single-mode distribution for any $\beta$ when $N_1 = N_2 = 0.5N$. When $N_1 = 0.5N + \delta n$ ($\delta n \ll N$), $P(m)$ is a single-mode distribution for $\beta < \beta_0$ and for $\beta > \beta_0$ a two-mode distribution (see Fig. 2b) without the peak at $m = 0$. When $N_1$ is noticeably different from $0.5N$ ($N_1 = 0.5N + \delta n$, $\delta n \sim N$), $P(m)$ is a single-mode distribution for $\beta < \beta_0$ and a three-mode distribution for $\beta > \beta_0$ (see Fig. 2c) with peaks emerging at points $m = \pm(N_1 - N_2)$ and the peak at $m = 0$ becoming very low
.

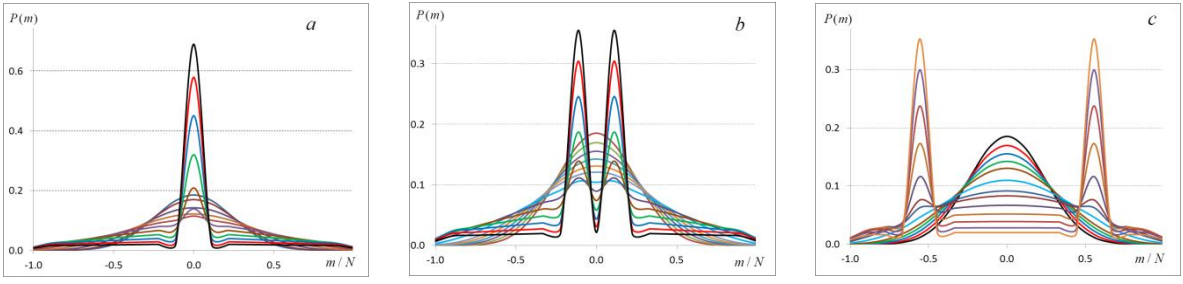

*Figure* 2. Temperature dependence of distribution $P(m)$ for different relations between $N_1$ and $N_2 = N - N_1$: a) $N_1 = 0.5N$; b) $N_1 = 0.55N$; c) $N_1 = 0.75N$.

The changes in curve $P(m)$ are shown in more detail in Figure 3. When $\beta > \beta_0$, $P(m)$ turns into a three-mode distribution: small peaks appear at the points corresponding to ground state $m = \pm(N_1 - N_2)$ (Fig. 3a). When $\beta = \beta_1$, these peaks become as high as the central peak (Fig. 3b). When $\beta = \beta_2$, the area under the lateral peaks becomes 1/2, and the height of the central peak becomes negligibly low (Fig. 3c).

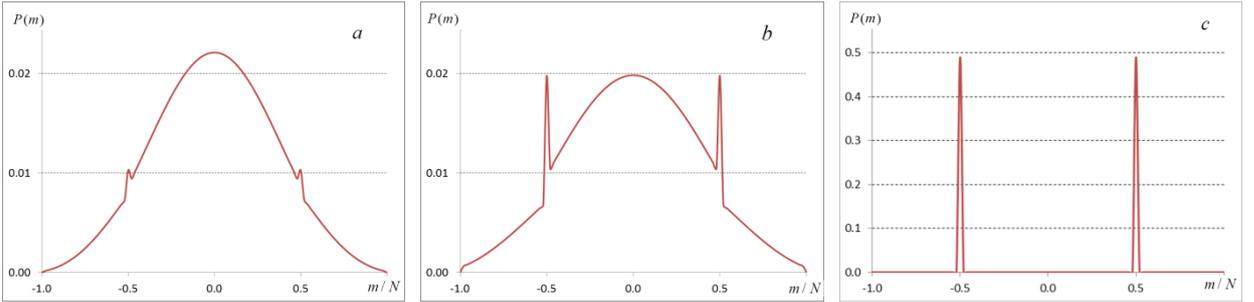

*Fi*gure 3. Curve $P(m)$ at $N_1 = 0.75N$, $N_2 = 0.25N$, $J_{11} = 1$, $J_{12} = -2$ and $J_{22} = 4$:
a) $\beta = 1.01\beta_0$, b) $\beta = \beta_1$, c) $\beta = \beta_2$.

### 7. Relation between magnetization and observation time

The relation between magnetization and observation time is highly dependent on the type of model under consideration. Here we confine ourselves to analysis of the model (13).

As was mentioned above, $P(m)$ is the probability of a thermodynamically stable system being in a state with magnetization $m$. From the physical point of view, it means that spontaneous magnetization $m$ will accrue in the system over a sufficiently long period of observation.

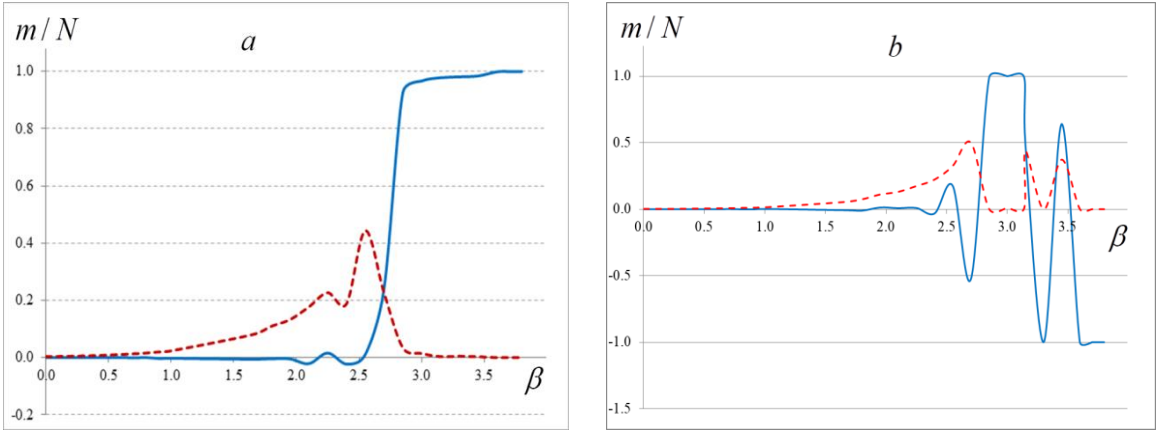

*Fi*gure 4. The solid-line curve corresponds to the dependence $m = m(\beta)$, the dotted-line curve is the magnetization dispersion. a) $10^5$ overturns per one spin; b) $4 \cdot 10^5$ overturns per one spin.

Clear that when $\beta \leq \beta_0$ and $P(m)$ is a single-mode distribution, there is no spontaneous magnetization (see Fig. 4). It can only be observed when $\beta > \beta_0$ and $P(m)$ becomes a multiple-mode distribution. Particularly, when there is a significant drop in temperature, we can observe spontaneous magnetization $|m| \to 1$ (Fig. 4a). This behavior is characteristic for relatively short observation times $\tau$ (when the system has not enough time to visit all states). If $\tau$ is long enough, we can observe the re-magnetization of the system when magnetization spontaneously changes sign (Fig. 3b). To evaluate the speed of these processes it is necessary to compare the time of the system being in a ground state (i.e. quantity $P_0 = \tau P(|m| = N)$) to the time of the system being in all other states $\tau(1 - 2P_0)$. For simplicity we assume that $\tau = 1$. The assumption allows us to omit this coefficient in further formulae.

First, let us determine temperature $\beta_2$, at which the system passes one half of the time in one of ground states and the other half in all other states, i.e. $2P_0 = 1 - 2P_0$. Since $P_0 = e^{-\beta E_0}/Z$ (where $Z$ is defined by (44)), the latter equality takes the form:

$$\frac{4}{\left(1 + e^{-2\beta}\right)^N + \left(1 - e^{-2\beta}\right)^N} = 1 \tag{48}$$

Quantity meeting (48) in the case of $N \gg 1$ has the form:

$$\beta_2 = \frac{1}{2} \operatorname{Ln} \frac{N}{\alpha} \tag{49}$$

where $\alpha = \operatorname{Ln}\left(2 + \sqrt{3}\right)$. (50)

Indeed, substituting $\beta = \beta_2$ in (48) we get the equation that is easily solvable with respect to $\alpha$:

$$\left(1+\frac{\alpha}{N}\right)^N + \left(1-\frac{\alpha}{N}\right)^N \simeq e^\alpha + e^{-\alpha} = 4$$

Note that the total number of states at $\beta = \beta_2$ is large enough. Nevertheless, the system spends half the time in two ground states ($|m| = N$), the other half of the time is spent wandering among many other states whose magnetization $|m| \neq N$. Additionally, comparison of (49) and (47) allows a conclusion that $\beta_2 > \beta_1 > \beta_0$.

Let us now consider how the re-magnetization time $\bar{P} = 1 - 2P_0$ changes when $\beta > \beta_2$. Let us introduce here dimensionless quantity $t$ characterizing departures from $\beta_2$:

$$t = \frac{\beta - \beta_2}{\beta_2} \tag{51}$$

As a result, we find that at temperatures $t > 0.5$ the re-magnetization time decreases exponentially:

$$\bar{P} = \frac{1}{2}\alpha^2 e^{-4\beta_2 t} \tag{52}$$

We have carried out a few numerical experiments to check the above expressions. We ran the Metropolis algorithm to calculate the mean and variance for each thermodynamic-equilibrium temperature and then to draw the graphs given in Fig. 4. Fig. 4a shows the measured values for a short observation time – $10^5$ overturns per a spin. As we can see, when $\beta > \beta_0$, the system demonstrates the presence of spontaneous magnetization: a relatively short observation time does not permit the re-magnetization of the system. Fig. 4b shows the measured values for a long observation time: $4 \cdot 10^5$ overturns per spin. Now we see that the observation time is long enough to allow the re-magnetization of the system.

To understand the process, in the Metropolis algorithm we measure the current value of magnetization at each time point at a given temperature: $2 \cdot 10^3$ overturns of spins were taken as unit time. The time dependences $m = m(\tau)$ at $\beta = 1.5\beta_2$ and $\beta = 2\beta_2$ are shown in Fig. 5. It is seen that in accordance with (52) the relative time of re-magnetization abruptly decreases with $\beta$ and curve $m = m(\tau)$ quickly degenerates into a meander.

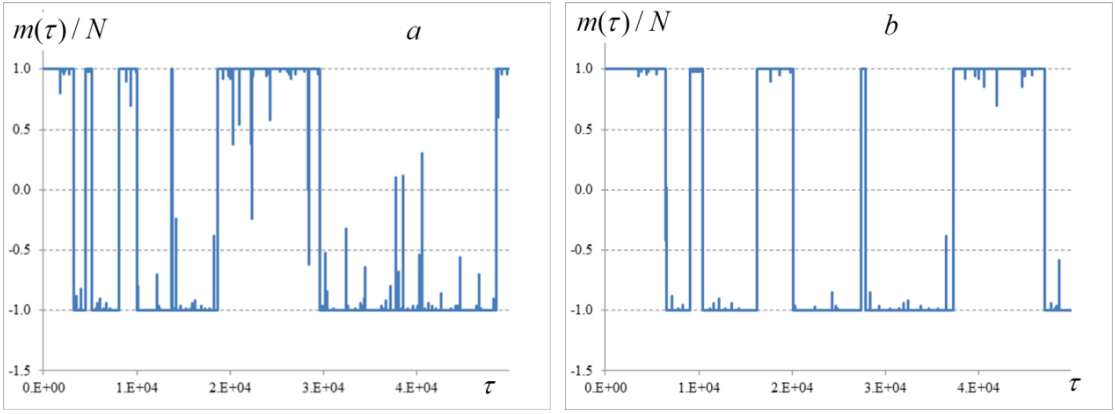

*Fi*gure 5. Time dependence of magnetization $m = m(\tau)$: a) $\beta = 1.5\beta_2$; b) $\beta = 2\beta_2$.

## 8. Discussing the results

The expressions we have obtained for the generalized density of states $D_N(E,m)$ are mostly referential because the expressions for the partition function of a single-dimensional model are well known. Correspondingly, almost all thermodynamic characteristics of a single-dimensional model are known. However, the knowledge of the generalized density of states $D_N(E,m)$ allows us to trace the distribution of system magnetization and show that non-zero spontaneous magnetization arises at a non-zero temperature.

The statement about the presence of spontaneous magnetization does not contradict the conventional result $\langle\langle m \rangle\rangle = 0$ where double brackets stand for the averaging over ensemble [1] - [3]. Indeed, it was shown above that the magnitude of spontaneous magnetization depends on time $m = m(\tau)$. This is due to the fact that the barrier between two ground states of the system is so small that the system passes quite easily from the vicinity of one ground state with magnetization $m \sim N$ to the vicinity of the other ground state with magnetization $m \sim -N$ and vice versa. We can conclude that given a long observation time, the time averaged magnitude of spontaneous magnetization $\langle m(\tau) \rangle$ is also zero, which is in full agreement with the ergodic theorem $\langle m(\tau) \rangle = \langle\langle m \rangle\rangle = 0$. We should only note that transitions between states $m \sim N$ and $m \sim -N$ are random and, therefore, the graphs shown in Fig. 4 and Fig.5 change from experiment to experiment.

The smallness of the energy barrier between the ground states is determined by the shape of the distribution $D_N(E,m)$ (see Fig. 6). In a single-dimensional chain (Fig. 6a) there are system configurations whose energy is close to the ground state $E = E_0 + 4$ and magnetization is continuously distributed in the interval $-N < m < N$. As a result, the system can pass from one ground state to another even at low temperatures. The picture changes in a two-dimensional model (Fig. 6b): there is no configuration whose energy is close to the ground state and magnetization is small. In this case, in order to move from one ground state to another, the system needs to overcome an energy barrier with height $\Delta E = 2\sqrt{N}$. This fact determines the stability of the ground state in the two-dimensional Ising model and occurrence of spontaneous magnetization.

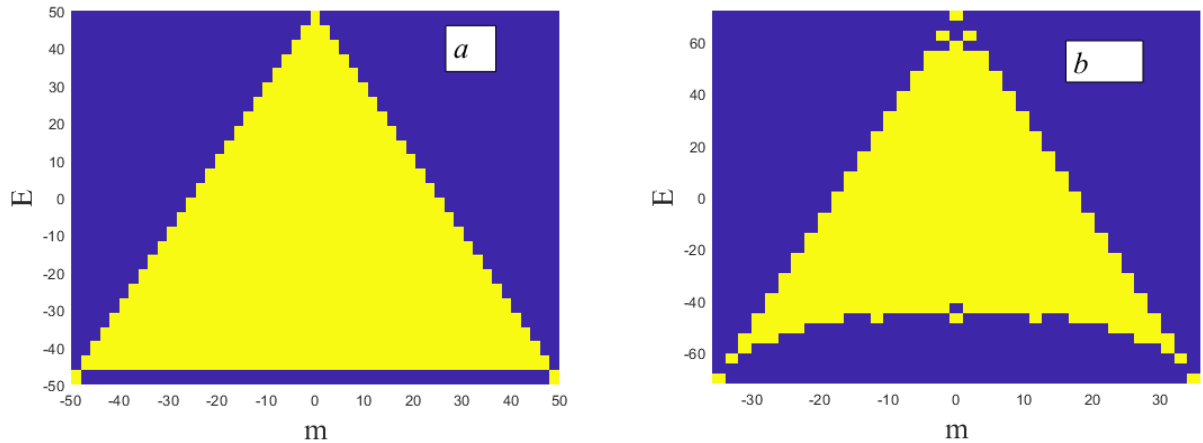

*Fig*ure 6. Comparison of the range of acceptable energy values $E$ and magnetization $m$ for (a) 1D Ising model and (b) 2D Ising model ($N = 50$). The blue indicates the areas where there are no state $D_N(E,m) = 0$, the yellow indicates the areas where the density of states is not zero ($D_N(E,m) > 0$). Both grids have periodic boundary conditions and ferromagnetic interactions.

**Appendix A**

The model characterized by energy (4) has a doubly degenerate ground state with energy $E = E_0$, where

$$E_0 = -N + 1 \qquad (A.1)$$

One of the ground states corresponds to the configuration with magnetization $m = N$ (all spins have upward direction: $s_i = 1$, $i = 1, 2, \ldots, N$), the other is characterized by magnetization $m = -N$ (all spins have downward direction: $s_i = -1$, $i = 1, 2, \ldots, N$).

Let us describe the calculation procedure. Let the system be in one of the ground states, e.g. all spins have upward direction. Let us mentally reverse the spins in a section of the chain. As a result, the system passes in the state with energy $E = E_0 + 2$ (the energy of the system increases by two because only the sign of the spin interaction at the boundaries of the section changes). If we repeat this operation $k$ times, we obtain a configuration with energy:

$$E = E_0 + 2k, \qquad k = 0, 1, \ldots, (N-1) \qquad (A.2)$$

Since there are $(N-1)$ interconnections between $N$ spins, the number of states with energy $E = E_0 + 2k$ for $k$ segments can be found as the number of $k$-combinations from a set of $(N-1)$ items:

$$D(E) = 2 \binom{N-1}{k} \qquad (A.3)$$

The number two appears in this expression because when reversing of spins in sections we can overturn either "right" or "left" part of the chain.

Let us now determine the generalized density of states $D_N(E, m)$ of a free chain. This sort of system has two ground states with energy $E_0 = -N + 1$, that is:

$$D_N(E_0, m) = \begin{cases} 1, & |m| = N \\ 0, & |m| \neq N \end{cases} \qquad (A.4)$$

Let us define the generalized density of states with higher energies $E > E_0$ ($k \geq 1$). By making $k$ section and overturns we obtain states with energy $E = E_0 + 2k$, the number $D(E)$ of which is determined by (A.3). Among these states there are a few configurations $D_N(E, m)$ in which $n$ spins look downwards and $N - n$ spins look upwards, i.e. the number of states with magnetization $m = N - 2n$. In this event the expression for $D_N(E, m)$ depends on the parity of the number of sections ($k$). To have the uniformity of further expressions, let us introduce here the notation:

$$r = \text{Int}\left(\frac{k+1}{2}\right) \qquad (A.5)$$

a) When $k$ is odd, there is an even number ($2r = k + 1$) of domains: $r$ domains with downwardly directed spins and $r$ domains with upwardly directed spins. Let downwardly directed domains contain $n$ spins, and upwardly directed domains hold $N - n$ spins. There are ($n - 1$) boundaries between $n$ spins in which we can insert ($r - 1$) sections. The number of possible configurations is found as the number of ($r - 1$) combination from a set of ($n - 1$) items. The number of configurations for ($N - n$) downwardly directed spins is calculated in a similar manner. As a result, the number of states with energy $E = E_0 + 2k$ and magnetization $m = N - 2n$ is described by the expression:

$$D_N(E, m) = 2\binom{n-1}{r-1}\binom{N-n-1}{r-1} \qquad (A.6)$$

where $r = \text{Int}[(k+1)/2] = (k+1)/2$, $k = (E + N - 1)/2$, $n = (N - m)/2$.

b) When $k$ is even, the operation with sectioning and spin overturns results in formation of $k + 1$ domains. Since the number of domains ($2r + 1 = k + 1$) is odd, two situations are possible: we have either r domains of n downwardly directed spins and (r + 1) domains of (N − n) of upwardly directed spins, or vice versa. Conducting reasoning similar to the case of odd $k$, we get the following expression for the density of states:

$$D_N(E, m) = \binom{n-1}{r}\binom{N-n-1}{r-1} + \binom{n-1}{r-1}\binom{N-n-1}{r} \qquad (A.7)$$

or, finally

$$D_N(E, m) = 2\frac{N-k}{k}\binom{n-1}{r-1}\binom{N-n-1}{r-1} \qquad (A.8)$$

where $r = \text{Int}[(k+1)/2] = k/2$, $k = (E + N - 1)/2$, $n = (N - m)/2$.

Combining (A.6) and (A.7), we get expression (8).

**Appendix B**

Let us consider a spin chain whose energy is determined by (13). First, we write down obvious expressions. The system has a doubly degenerate ground state. The configurations providing the ground state have magnetization $m = N$ and $m = -N$. The energy of the ground state ($E = E_0$) is

$$E_0 = -N \tag{B.1}$$

Correspondingly, we have

$$D_N(E_0, m) = \begin{cases} 1, & |m| = N \\ 0, & |m| \neq N \end{cases} \tag{B.2}$$

Let us consider the generalized density of states with higher energy values $E > E_0$. Clear that we have to make $2k$ sections and $k$ overturns of spins in them to pass from the ground state to the state with energy $E > E_0$. As a result, we arrive at the state with energy

$$E = E_0 + 4k, \quad k = 0, 1, ..., k_{max}, \quad k_{max} = \text{Int}(N/2) \tag{B.3}$$

The number of states with energy $E = E_0 + 4k$ and magnetization $m$ is determined as

$$D_N(E, m) = \frac{N}{k}\binom{n-1}{k-1}\binom{N-n-1}{k-1}, \quad k = (E+N)/4, \quad n = (N-m)/2. \tag{B.4}$$

The way we come to expression (B.4) can be explained by the following reasoning. Let us make the first of $2k$ sections between the first and second spins. As a result, we come to the problem of a spin chain with free boundary conditions and $2k-1$ sections that complies with expression (A.8). It seems that we can make the first section between any spins and get $N$ possible configurations. However, it is easy to see that when the first section is shifted by $(N/2k)+1$ positions, the pattern repeats itself, i.e there are $N/2k$ non-repeating combinations. Thus, making substitution $r \to k$ and multiplication by $N/2k$ in (A.8), we obtain expression (B.4).

**Appendix C**

Let us consider a free chain whose energy is defined by (22) – (23). It follows from (22) that the energy of the system is $E = E_1 + E_2 - J_{12}$ when $s_{N_1} s_{N_1+1} = 1$, or $E = E_1 + E_2 + J_{12}$ when $s_{N_1} s_{N_1+1} = -1$. It is, therefore, necessary to determine the probability of these spins $s_{N_1}$ and $s_{N_1+1}$ being turned in the same direction or in opposite directions.

Given the first chain, the probability is dependent on the parity of the number of cuts $k_1$. If $k_1$ is an odd number, the spins at the ends of the chain are oppositely directed $s_1 s_{N_1} = -1$. At the same time, for each spin configuration it is possible to uniquely set a mirror configuration ($s_i \to s_{N_1-i+1}$) with the same $k_1$ and $m$. Hence with an odd number of $k_1$ one half of configurations has $s_{N_1} = 1$, the other $s_{N_1} = -1$. If $k_1$ is an even number, the number of configurations with $s_{N_1} = 1$ is equal to the number of configurations with $(r+1)$ upwardly directed domains and $r$ downwardly directed domains. Similarly, the number of configurations with $s_{N_1} = -1$ is equal to the number of configurations with $(r+1)$ downwardly directed domains and $r$ upwardly directed domains. The expressions for these cases are obtained in Appendix A (see the summands in (A.7)). As we see, the probability of the boundary spin $s_{N_1}$ looking upward is:

$$p_\uparrow(1) = \begin{cases} a_1, & k_1 \text{ is even} \\ 1/2, & k_1 \text{ is odd} \end{cases}, \quad \text{where} \quad a_1 = \frac{n_1 - k_1/2}{N_1 - k_1} \tag{C.1}$$

and the probability of the boundary spin $s_{N_1}$ looking downward is:

$$p_\downarrow(1) = \begin{cases} b_1, & k_1 \text{ is even} \\ 1/2, & k_1 \text{ is odd} \end{cases}, \quad \text{where} \quad b_1 = \frac{N_1 - n_1 - k_1/2}{N_1 - k_1} \tag{C.2}$$

Similar expressions can be written for spin $s_{N_1+1}$ by replacing index 1 by 2 in (C.1) – (C.2).

The probability that the energy of the united system is $E = E_1 + E_2 - J_{12}$ (spins $s_{N_1}$ and $s_{N_1+1}$ have the same direction) has the form:

$$P_{\uparrow\uparrow} = p_\uparrow(1) p_\uparrow(2) + p_\downarrow(1) p_\downarrow(2) \tag{C.3}$$

and the probability that the energy of the united system is $E = E_1 + E_2 + J_{12}$ (spins $s_{N_1}$ and $s_{N_1+1}$ have the opposite directions) is

$$P_{\uparrow\downarrow} = p_\uparrow(1) p_\downarrow(2) + p_\downarrow(1) p_\uparrow(2) \tag{C.4}$$

In view of the designations introduced in (C.1) and (C.2), we finally get (26) – (28).

**Appendix D**

Let us consider a model with periodic boundary conditions whose energy is determined by (31) – (32). Let us represent it as a sum of two chains with free boundary conditions.

a). If $k_1 \bmod 2 \neq k_2 \bmod 2$, i.e. the numbers of cuts $k_1$ and $k_2$ in the chains have opposite parity, then $s_1 s_N + s_{N_1} s_{N_1+1} = 0$ and the energy of the united system is $E = E_1 + E_2$.

b). If $k_1 \mod 2 = k_2 \mod 2$, then both interconnections between the chains are the same $s_1 s_N = s_{N_1} s_{N_1+1}$. As follows from (31), the energy of the system is $E = E_1 + E_2 - 2J_{12}$ when $s_1 s_N = s_{N_1} s_{N_1+1} = 1$, or $E = E_1 + E_2 + 2J_{12}$ when $s_1 s_N = s_{N_1} s_{N_1+1} = -1$. When $s_{N_1} s_{N_1+1} = 1$ and $s_{N_1} s_{N_1+1} = -1$ the probabilities are defined by expressions (C.3) and (C.4) correspondingly.

As a result, we obtain expressions (35) and (36).

**Appendix E**

Let us consider the case when the energy of the chain is set by the expression:

$$E = -J_{12} s_1 s_N - J_{11} \sum_{i=1}^{N-1} s_i s_{i+1} \tag{E.1}$$

where $J_{11} > 0$, and $J_{12}$ can take any value.

This system has $2k$ sections which form domain boundaries. The probability of a domain boundary being at the end of the chain ($s_1 s_N = -1$) is

$$p_{\uparrow\downarrow}(k) = \frac{2k}{N} \tag{E.2}$$

Similarly, the probability of $s_1 s_N = 1$ is

$$p_{\uparrow\uparrow}(k) = \frac{N - 2k}{N} \tag{E.3}$$

If $s_1 s_N = -1$, the energy of the system is

$$E = J_{11}(-N + 4k - 1) + J_{12} \tag{E.4}$$

If $s_1 s_N = 1$, the energy is

$$E = J_{11}(-N + 4k + 1) + J_{12} \tag{E.5}$$

The density of states can be expressed via the expression for a periodic-boundary-condition chain (16) in the following fashion:

$$D_N(E, m) = \sum_k \frac{N}{k} \binom{n-1}{k-1}\binom{N-n-1}{k-1}\left(p_{\uparrow\downarrow}(k)\delta_{E, J_{11}(-N+4k-1)+J_{12}} + p_{\uparrow\uparrow}(k)\delta_{E, J_{11}(-N+4k+1)+J_{12}}\right) \tag{E.6}$$

In the specific case of $J_{11} = 1$ and $J_{12} = -1$, expression (E.6) becomes noticeably simpler:

$$D_N(E, m) = p_{\uparrow\downarrow}(k+1) \frac{N}{k+1}\binom{n-1}{k}\binom{N-n-1}{k} + p_{\uparrow\uparrow}(k) \frac{N}{k}\binom{n-1}{k-1}\binom{N-n-1}{k-1} = \\ = \frac{2n(N-n) - kN}{n(N-n)} \cdot \binom{n}{k}\binom{N-n}{k}, \tag{E.7}$$

where $E = -N + 2 + 4k$.